\documentclass[twocolumn,prb,superscriptaddress]{revtex4}

\usepackage{epsfig,amsmath,amssymb,color}
\newcommand{\be}{\begin{equation}}
\newcommand{\ee}{\end{equation}}
\newcommand{\bea}{\begin{eqnarray}}
\newcommand{\eea}{\end{eqnarray}}
\newcommand{\up}{\uparrow}
\newcommand{\down}{\downarrow}
\def\nn{\nonumber\\}

\def\fr#1{(\ref{#1})}

\def\eps{\epsilon}

\def\ontop#1#2{\genfrac{}{}{0pt}{}{#1}{#2}}
\begin{document}

\title{Finite temperature dynamics of the Mott insulating Hubbard
  chain}
\author{Alberto Nocera}
\affiliation{Department of Physics and Astronomy, University of Tennessee, Knoxville, TN 37966, USA}
\affiliation{Materials Science and Technology Division, Oak Ridge National Laboratory, Oak Ridge, TN 37831, USA}
\author{Fabian H. L. Essler}
\affiliation{Rudolf Peierls Centre for Theoretical Physics, University of Oxford, Oxford OX1 3NP, United Kingdom}
\author{Adrian E. Feiguin}
\affiliation{Department of Physics, Northeastern University, Boston, Massachusetts 02115, USA}

\date{\today}
\begin{abstract}
We study the dynamical response of the half-filled one-dimensional(1d)
Hubbard model for a range of interaction strengths $U$ and
temperatures $T$ by a combination of numerical and analytical
techniques. Using time-dependent density matrix renormalization group (tDMRG) computations we find that the single-particle
spectral function undergoes a crossover to a spin-incoherent Luttinger
liquid regime at temperatures $T \sim J=4t^2/U$ for sufficiently large
$U > 4t$. At smaller values of $U$ and elevated temperatures the
spectral function is found to exhibit two thermally broadened bands of
excitations, reminiscent of what is found in the Hubbard-I approximation. 
The dynamical density-density response function is shown to exhibit
a finite temperature resonance at low frequencies inside the Mott gap,
with a physical origin similar to the Villain mode in gapped quantum spin
chains. We complement our numerical computations by developing an
analytic strong-coupling approach to the low-temperature dynamics in
the spin-incoherent regime.
\end{abstract}

\maketitle

\section{Introduction}

The physics of one-dimensional metals can generally be described in
terms of Luttinger liquid theory. In a Luttinger liquid (LL)
\cite{Haldane1981,Gogolin,GiamarchiBook}, the natural excitations are
collective density fluctuations, that carry either spin (``spinons''),
or charge (``holons''). This leads to the spin-charge separation
picture, in which a fermion injected into the system breaks down into
excitations carrying different quantum numbers, each with a
characteristic energy scale and velocity (one for the charge, one for
the spin). A key paradigm for this kind of behavior is provided by the
one-dimensional Hubbard model:
\begin{equation}
H=-t \sum_{i=1,\sigma}^L \left(c^\dagger_{i\sigma} c_{i+1\sigma}+\mathrm{h.c.}\right)
+ U \sum_{i=1}^L \big(n_{i\uparrow}-\frac{1}{2}\big)\big(n_{i\downarrow}-\frac{1}{2}\big).
\label{Hubbard}
\end{equation}
Here, $c^\dagger_{i\sigma}$ creates an electron of spin $\sigma$ on the
$i^{\rm th}$ site along a chain of length $L$. The Coulomb repulsion is parametrized by $U$, and we take the inter-atomic distance as unity. We express all energies in units of the hopping parameter $t$.


A remarkable aspect of this model is that it is
integrable: it contains an extensive number of local integrals of
motion that allow one to exactly solve it with the Bethe Ansatz
\cite{EsslerBook}. 
Whereas the low-energy physics of the 1d Hubbard
model below half filling is described in terms of LL
theory\cite{frahm90}, at half-filling (density of particles $n=1$, or
number of particles $N=L$) the model has a
Mott-insulating ground state, with a charge gap that
grows exponentially with $U$ for weak interactions. The spin
excitations, however, remain gapless and the system exhibits
algebraically decaying antiferromagnetic spin-spin correlations.  
Mott insulators defy conventional paradigms, since the rigid band
picture underlying the physics of semiconductors does not
apply \cite{Eskes1991,PhillipsRMP2010}: in strongly interacting
systems, the ``bands'' change with doping, giving rise to a complex
phenomenology that includes hole pockets, Fermi arcs and kinks
\cite{Damascelli2003,Meng2009}.  

The zero temperature dynamical properties of the Hubbard model have
been studied in great detail by a variety of analytic and numerical
methods both in the
metallic\cite{frahm90,Sorella1991,xiang1992charge,Penc1997,penc97,jeckelmann00,benthien04,carmelo04,carmelo06,carmelo08,Kohno2010,SIG10a,pereira10,essler10,SIG10b,SIG12,seabra14,essler15,tiegel16,veness16,Yang2016}
and the Mott insulating\cite{voit98,doyon02,lante09,pereira12} phases.
The finite temperature dynamics is less well understood.
The single-particle spectral function below
  half-filling has been previously studied by
quantum Monte Carlo\cite{Abendschein2006} and density matrix
renormalization group (DMRG)\cite{Feiguin2011} methods.
The half-filled case was considered in Ref.~\onlinecite{maekawa05} in the
small to intermediate temperature regime, $T\leq t$, using QMC with
maximum-entropy analytic continuation procedures. The low-temperature
regime for small Mott gaps $T\alt\Delta\ll t$ was analyzed by field
theory methods in Ref.~\onlinecite{essler03}.
As a result of spin-charge separation one-dimensional metals and Mott
insulators can display very unusual behavior at finite temperatures.
An example is the so-called ``spin-incoherent'' Luttinger liquid regime
\cite{Matveev2004,Fiete2004,Cheianov2004,Cheianov2005,Fiete2007b,Halperin2007},
which occurs in the metallic case if the spinon bandwidth is much
smaller than the holon bandwidth, which corresponds to $U \gg t$ in
case of the Hubbard chain. In this regime a small temperature makes
the spin degrees of freedom completely incoherent, while the charge
degrees of freedom remain close to the ground state. In this
situation, excitations effectively behave as ``spinless fermions''
and this has dramatic effects on the spectral functions
\cite{Feiguin2009d, Soltanieh-ha2014}.  

The finite temperature dynamics of quasi-one-dimensional Mott
insulators has been investigated in a number of cases. Spin-charge
separation was observed in photoemission experiments on the chain
cuprates SrCuO$_2$\cite{Kim1996,Kim1997,Kim2006} and
Sr$_2$CuO$_3$\cite{kidd08}. In these materials the characteristic
energy scales for spin and charge degrees of freedom are very large,
so that the achievable temperatures are always small compared to the
Mott gap. The experimental findings could not be accounted for by a
simple $t$-$J$ model at zero temperature.
ARPES
measurements on the one dimensional Mott insulator
Na$_{0.96}$V$_2$O$_5$ show that the spectral density of the lower
Hubbard band is strongly dependent on the temperature\cite{maekawa99}. Simple
broadening and charging effects could not 
explain the dramatic spectral weight redistribution as a function of
temperature, which was attribuited to strong correlation effects in
the material. Exact diagonalization studies of a $t-J$ model presented
discrepancies with the experimental results. In particular, while
theory and experiment agreed well for small momentum transfers and at
low energy, a long tail of excitations at high energies for
intermediate-to-large momentum transfers (up to the Brillouin zone
boundary) was observed and not accounted for by the theory. 
Some of the experimentally observed effects are expected
to be due to the fact that the appropriate effective Hamiltonians for
the various materials will contain longer-range hoppings and
interactions, while others will be due to finite temperatures. To get
a qualitative understanding of the latter it is clearly useful to
investigate the dynamics of the one dimensional Hubbard model in a
range of temperatures.


Experimentally probing the high-temperature dynamics
at a fixed density has remained a challenge, since the Mott gap is of
the order, or higher than the melting point of some
materials. However, several on-going efforts in the cold-atom
community are focused on studying the excitations of
spinful fermionic systems\cite{Sagi2015}, particularly in
one dimension\cite{White2017}.  

The paper is structured as follows: In Section \ref{method} we briefly describe the computational methods; in Section \ref{results} we present the results of our finite-temperature simulations. Section \ref{sec:SC} describes an analytical approach for the spin-incoherent regime. We finally close with a summary and discussion.

\section{Numerical approach}\label{method}

  In this work we would like to explore the entire temperature regime at half-filling. For this purpose we resort to finite temperature time-dependent DMRG calculations (tDMRG)\cite{White2004,Daley2004,Feiguin2011vietri,Feiguin2013a}. Even though the method has been extensively discussed in the literature \cite{Feiguin2005a}, particularly in a recent review \cite{Feiguin2013a}, we proceed to describe it in a condensed form.

The calculation relies on ideas from thermo field dynamics.\cite{Takahashi1975,UmezawaBook,Matsumoto1986,Suzuki1985,Suzuki1985b,Barnett1988,Barnett1985} This construction allows one to
represent a mixed state of a quantum system as a pure state in an enlarged
Hilbert space.
Consider the energy eigenstates of the system in
question $\{n\}$, described by a Hamiltonian $H$, and introduce an auxiliary set
of fictitious states $\{\tilde{n}\}$ in one-to-one correspondence with $\{n\}$.
We can then define the unnormalized pure quantum state,
\begin{equation}
| \psi(\beta) \rangle = e^{-\beta H/2}| \psi(0) \rangle = \sum_n
e^{-\beta E_n/2} |n \tilde{n}\rangle \label{thermoa}
\end{equation}
where $\tilde{n}$ is a copy of $n$ in the auxiliary Hilbert space, $\beta=1/T$ is the
inverse temperature, and $|\psi(0)\rangle=\sum_n{|n\tilde{n}\rangle}$ is our
thermal vacuum.
Then the exact thermodynamic average of an operator $\hat O$ (acting
only on the real states), is given by
\begin{equation}
\langle \hat O \rangle = Z(\beta)^{-1} \langle \psi(\beta) | \hat O | \psi(\beta) \rangle.
\label{average}
\end{equation}
Here, the partition function is the norm of the thermal state $Z(\beta)=\langle
\psi(\beta) | \psi(\beta) \rangle$.
Hence, a thermodynamic average reduces to
a conventional expectation value in a pure quantum
state.

At $\beta=0$, the state $|\psi(0)\rangle$ is the maximally entangled state between
the real system and the fictitious system. We can see that this is independent of
the representation, and we can choose any arbitrary basis.

\begin{figure}
\centering
\includegraphics[width=0.48\textwidth]{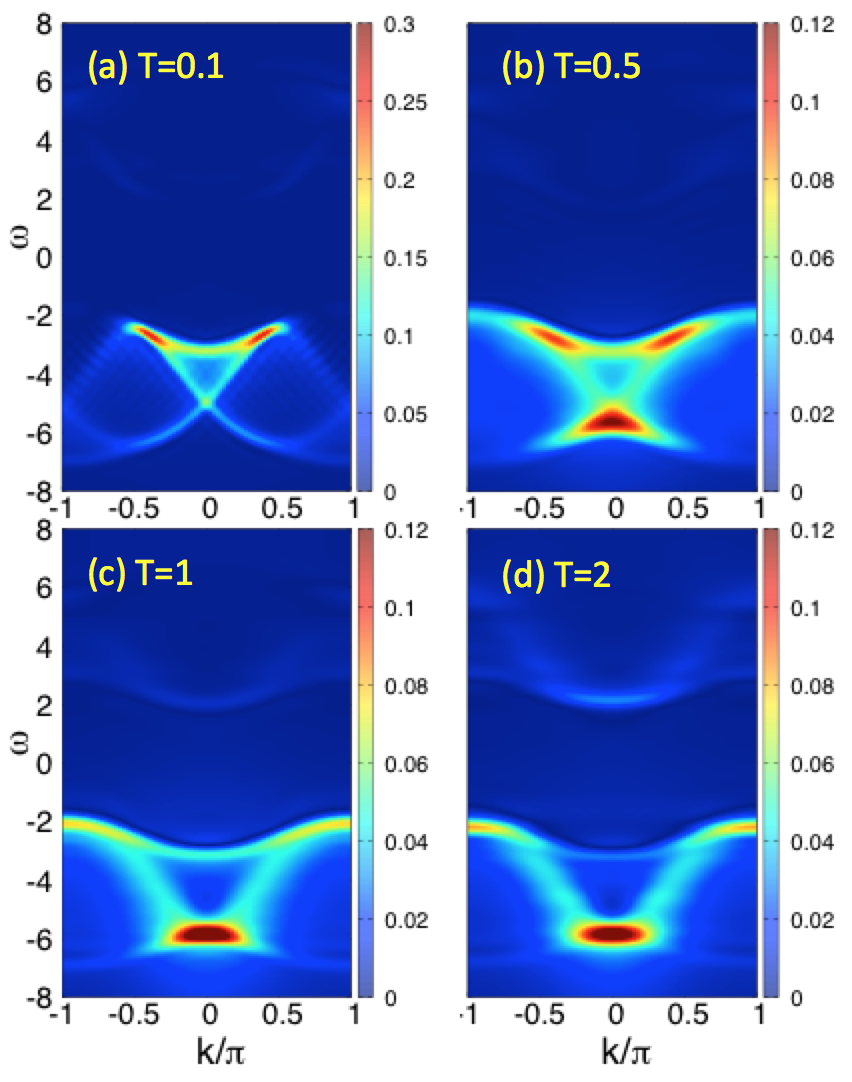}
\caption{Photoemission spectrum for the 1d Hubbard model with $U/t=8$ calculated with tDMRG at four different temperatures.}
\label{fig:U8}
\end{figure}

The most technical aspect concerns the choice of initial state at $\beta=0$. One can pick to work in either the grand canonical, or the canonical ensemble \cite{Feiguin2011,Nocera2016,Barthel2016}, and this will determine the method to initialize the simulations. 
In order to work in the canonical ensemble we need to start from a thermal vacuum where the physical states $|n\rangle$ and their copies $|\tilde{n}\rangle$ have each a fixed number of particles.
This requires one to construct a state that is a sum of all possible states of charge and spin, with the constraint that the total number of particles on the chain has to be equal to $N$, and that the charge state of the ancillas is an exact copy of the charge state of the physical chain. In order to generate this state, we use the conventional ground-state DMRG algorithm with a very peculiar Hamiltonian (we can call it the ``entangler Hamiltonian''):
\begin{equation}
H_\mathrm{ent} = -\sum\limits_{i\neq j \\ \sigma} \left(\Delta^\dagger_{i\sigma} \Delta_{j\sigma}+\mathrm{h.c.}\right).
\label{entangler}
\end{equation}
The operator $\Delta^{\dagger}$($\Delta$) is given by 
\begin{equation}
\Delta^\dagger_{i\sigma} = c^\dagger_{i\sigma}\tilde c^\dagger_{i\sigma},
\end{equation}
where the ``tilde'' operators act on the ancillas on site $i$.
The ground state of this Hamiltonian is precisely the equal superposition of all possible configurations of $N$ ``physical site-ancilla'' pairs on $L$ sites.

The Green's functions at time $t$ and inverse temperature $\beta$ is obtained as
\begin{equation}
G(x-x_0,t,\beta)=\langle \psi(\beta)|e^{i\hat{H}t}\hat O^{\dagger}(x)e^{-i\hat{H}t}\hat O(x_0)|\psi(\beta)\rangle,
\label{gf}
\end{equation}
where the generic operators of interest $\hat O(x)$,$\hat O^\dagger(x)$ act on the system at site $x$. The time evolution is dictated by the Hamiltonian $\hat{H}=H-\tilde{H}$; $H$ governs the physics of the actual physical chain, not including the ancillas, while $\tilde{H}$ is an exact copy of $H$ acting on the ancillas \cite{Karrasch2013}.

The calculation proceeds as follows: First, we evolve the maximally entangled state in imaginary time to the desired value of $\beta$ (measured in units of the hopping $t$). Then, an operator $\hat O(x_0=L/2)$ is applied in the center of the chain. The resulting state is evolved in real-time, and at every step
we measure the overlap with the state $\hat O(x)e^{-i(H-\tilde{H})t}|\psi(\beta)\rangle$.
We obtain the desired Green's function in frequency and momentum by Fourier transforming the results in real space and time.
In this work we use a third order Suzuki-Trotter decomposition with a typical time-step $\tau_\beta=0.05$ and $\tau=0.05$ for the real-time and imaginary-time parts of the simulation, and keeping 800 DMRG states, enough to maintain the truncation error below $10^{-5}$. The chain length is fixed to $L=40$ sites, and the real-time window has a range $t_\mathrm{max}=15$.

\begin{figure}
\centering
\includegraphics[width=0.46\textwidth]{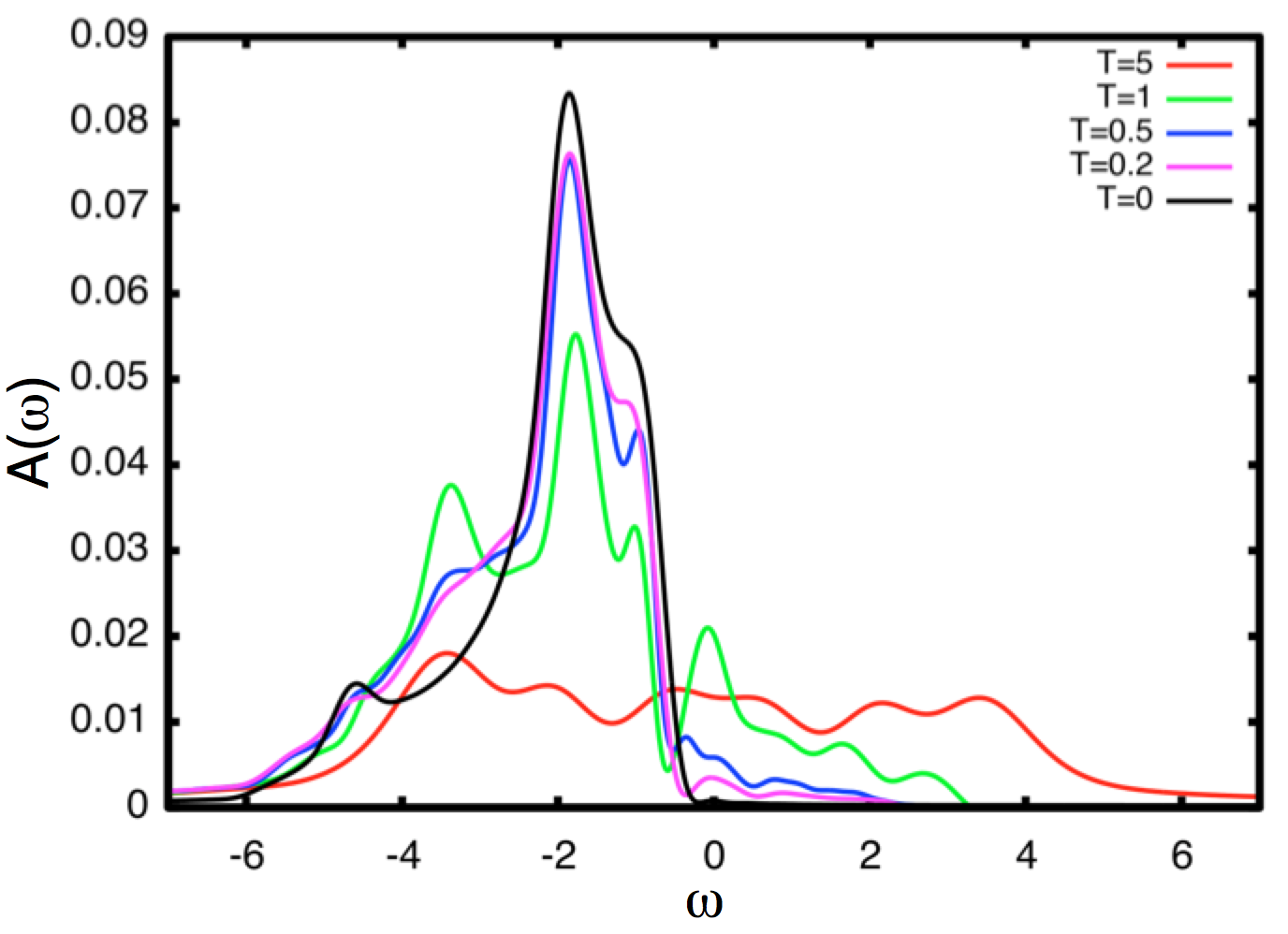}
\caption{Integrated photoemission spectrum (occupied fraction of the total density of states) for the 1d Hubbard model at half-filling with $U/t=4$ obtained with tDMRG at different temperatures. }
\label{fig:U4}
\end{figure}

\section{Results}\label{results}
\subsection{Single-particle spectral function}

\begin{figure}
\centering
\includegraphics[width=0.48\textwidth]{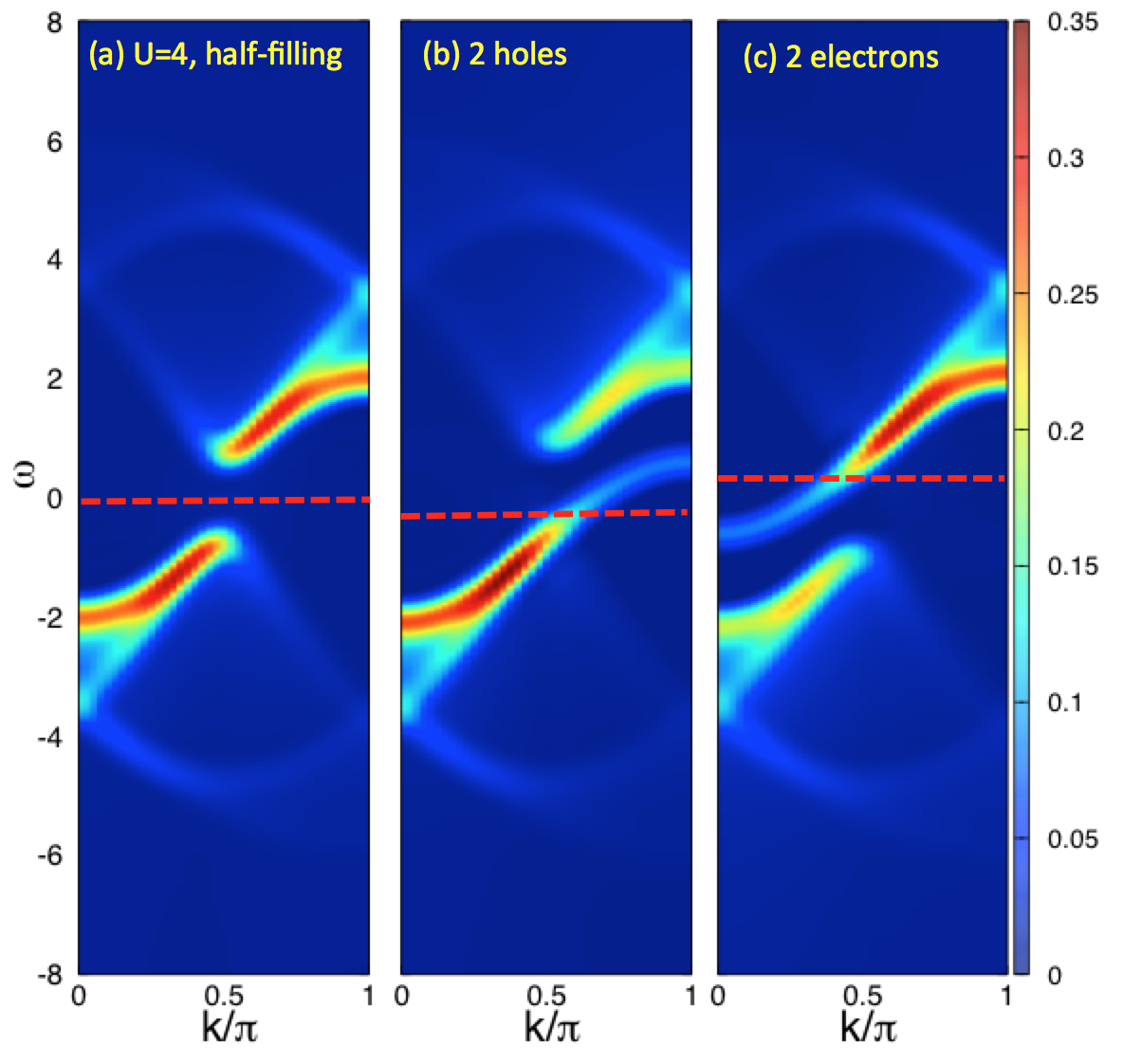}
\caption{Zero temperature spectral function of the 1d Hubbard model with $U/t=4$ calculated with tDMRG at three different densities: (a) half-filling, (b) 2 hole doping and (c) 2 additional electrons. The dashed line indicates the position of the chemical potential. The mere addition/subtraction of a single particle immediately destroys the Mott gap. }
\label{fig:T0}
\end{figure}

 The electronic spectrum of the Hubbard model can be qualitatively understood in the $U \rightarrow \infty$ limit\cite{Ogata1990,Ogata1991}, where spin-charge separation is exact. In this regime, all eigenstates factorize into a product of a fermionic wave function and a spin wave function. This leads to a simple and elegant description\cite{Suzuura1997,Brunner2000}: assuming that the dispersion of holons is given by $\epsilon_h(q_h)=-2t\cos q_h$, and the one for spinons by $\epsilon_s(q_s) = J \cos q_h$, with $J=4t^2/U$, one can construct all possible energies with momentum $k$ as $\epsilon(k)=\epsilon_h(q_h) + \epsilon_s(q_s)$, with $k=q_h+q_s$. Clearly, this construction will yield a continuum of energies with momentum $k$: the Fermi-liquid description breaks down, and there is no fermionic quasi-particle in the Landau sense. The elementary excitations of the Hubbard model can then be summarized as spinons, and holons below the Fermi energy. Holons are defined between $q_h = \pm 2k_F$, and removing a particle from the system corresponds to creating deconfined spinon-holon pairs. The excitations for adding a particle lead to the creation of spinon-antiholon pairs above the Fermi energy. Antiholons are defined in the intervals $[-\pi,-2k_F]$ and $[2k_F,\pi]$ and usually refer to {\it empty} states
as shown in Fig.~\ref{fig:T0}(a). 
Notice that this picture also applies to the upper Hubbard band, but in this case we name the particle excitations as ``doublons''. 

In Fig.\ref{fig:U8} we show the photoemission part of the spectral function for $U/t=8$ and four different temperatures, obtained with a choice of operators $\hat{O} = c$ (the annihiliation operator);  this corresponds to the {\it occupied} electronic states. At low temperatures (upper panels) we can clearly resolve both holon and spinon bands. Even at small temperatures relative to the gap, we see that there is a finite occupation of the upper Hubbard band. Most remarkably, for increasing temperatures, we find spectral weight leaking into the gap. As doublons are created, holes are left behind in the lower Hubbard band and thermally excited quasiparticles can now occupy these states. This gives rise to weight in the gap due to anomalous spectral transfer\cite{Eskes1991}.
Near the Fermi level, high energy electrons can now occupy states that are attributed to antiholons, which at zero temperature are always {\it empty} states. The antiholon-spinon continuum leaks into the gap near the Mott transition \cite{Kohno2012}, as we have seen above in Fig.~\ref{fig:T0}. 
The thermal excitations can be traced back to the antiholon-spinon continuum that appears in the gap, and their particle-hole conjugates in the upper Hubbard band.
This leads to a melting of the gap which is even more dramatic at smaller values of $U/t$, as seen in Fig.\ref{fig:U4} where we display the photoemission spectrum integrated over momentum: for $U/t=4$ the gap is gone altogether at small temperatures.  
In addition, while the Mott gap is melting, spectral weight is redistributed to large
and negative energies, in qualitative agreement with the experimental results in
Ref.~\onlinecite{maekawa99}.
The origin of the in-gap states can be traced back to the zero
temperature spectra of the doped Mott insulator. For illustration, in
Fig.~\ref{fig:T0} we show the zero-temperature spectrum of the chain
with 2 holes and 2 electrons. The states below (above) the Fermi level 
are occupied (empty) and the spinon-antiholon branches are responsible for the leakage of spectral weight into the gap. We clearly see that the mere
addition/subtraction of a particle can completely melt the Mott gap
{\it in the finite system} considered. In the thermodynamic limit, a
finite density of holes would be required to make this effect
observable. However, a finite temperature induces similar effects. 
An important issue to address is the behavior of the spin excitations with temperature \cite{Abendschein2006a}. As we see in our results for $U/t=8$, the ``two-branch'' spectrum characteristic of a Luttinger Liquid survives to temperatures of the order of $T \sim J$. 
That is when the crossover to the spin-incoherent regime takes place: at higher temperatures the spinons are effectively thermalized, and the spectrum resembles that of spinless fermions with a single branch $\epsilon(k)=-2t\cos(k)$. This is remarkable considering that the excitations are completely determined by the Hamiltonian, and do not change with temperature: what changes is the distribution of spectral weight. We need to recall that the electronic Green's function is a convolution of the charge and spin Green's functions. At large values of $U$ the spinon band is less dispersive than the charge excitations and the spectral weights will respond differently for each kind of excitations. The charge will behave as though in the ground state, but the redistribution of spectral weight in the spin Green's function will lead to a band-like feature that shifts {\it in momentum}. We refer the reader to Refs.\onlinecite{Feiguin2009d, Soltanieh-ha2014} for a detailed description of the phenomenon. 
Notice that our results are in qualitative agreement with those obtained in Ref.~\onlinecite{maekawa05}, where the authors used QMC with maximum-entropy analytic continuation procedures in the low temperature regime. 

\begin{figure}
\centering
\includegraphics[width=0.48\textwidth]{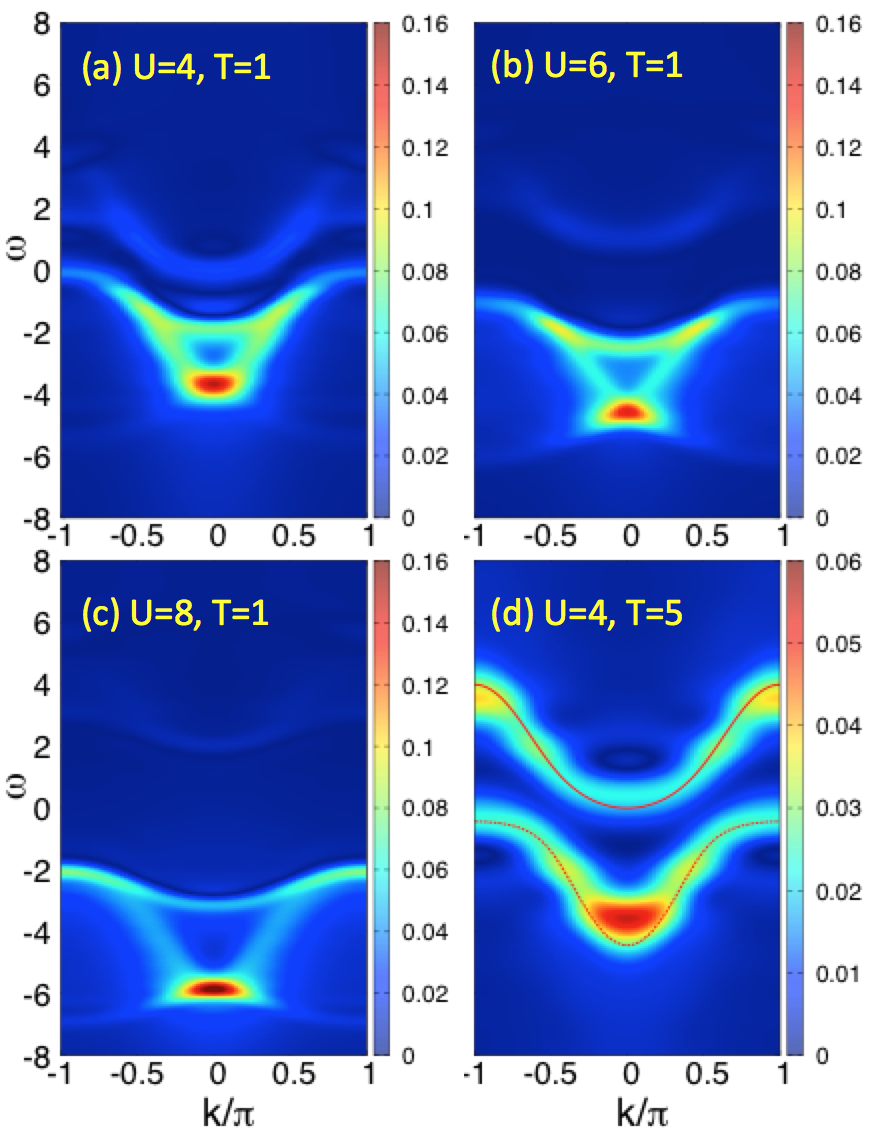}
\caption{(a-c) Photoemission spectrum for the 1d Hubbard model with $U/t=4,6,8$ and $T=1$ calculated with tDMRG. In panel (d) we show results for $U=4$ and $T=5$. We have included the dispersion obtained from the Hubbard I approximation (see text), with a renormalized $U_\mathrm{eff}$.}
\label{fig:U4-6}
\end{figure}

For smaller values of $U/t$ we do not find spin-incoherent behavior. In this case the spin and charge dispersions have broad bandwidths, and all degrees of freedom will get similarly excited. At temperatures larger than the hopping $T>t$, we find completely incoherent upper and lower Hubbard bands. This regime can be qualitatively described in terms of a Hubbard-I mean-field approximation \cite{Dorneich2000,Groeber2000}, with interacting doublon and holon excitations that ignore the magnetic correlations. In this case the two bands are given by:
\begin{equation}
E_\pm(k) = \epsilon(k)\pm \sqrt{\epsilon(k)^2 + \left(\frac{U_\mathrm{eff}}{2}\right)^2},
\label{HubI}
\end{equation}
where $\epsilon(k)=-2t_\mathrm{eff}\cos(k)$ is once again the non-interacting dispersion but with a renormalized hopping. In the original formulation, the excitations are expected to be twice as heavy ($t_\mathrm{eff}=t/2$) because, due to the spin-incoherent background, a hole or doublon have half the probability to hop to a neighboring site. This approach works very well in 2D at high temperatures, but as we see in Fig.\ref{fig:U4-6}(d), in one-dimension the bandwidth remains unaffected, with $t_\mathrm{eff}=t$ and an effective $U_\mathrm{eff} \leq U$. This is attributed once again to spin-charge separation: in one-dimension the holon and doublon excitations can be assumed to be spinless quasi-particles with hopping $t$, regardless of the spin background. 

\begin{figure}
  \centering
  \includegraphics[width=0.35\textwidth]{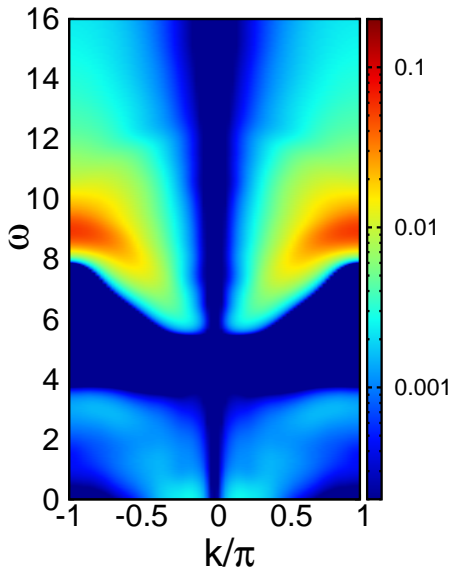}
  \caption{Charge dynamical structure factor for $U=8$, $T=1$. The finite spectral weight at low frequencies is a result of intraband transitions. Notice that color density is in a logarithmic scale.
}
  \label{fig:density}
\end{figure}
\subsection{Villain-like mode}
An interesting feature in the dynamical response of gapped
many-particle systems are finite temperature resonances at low
frequencies. The paradigm for this kind of behavior is the so-called
Villain mode \cite{Villain1975} in the spin-1/2 Heisenberg chain.
The mechanism underlying this feature is straightforward: at
temperatures of the order of the gap, a thermal population of spin
excitations will result in a dynamical spin response at low energies 
$\omega \sim 0$ due to transitions between thermally occupied states.
The weight of these contributions to the dynamical spin structure
factor increases with temperature at low $T<\Delta$, where $\Delta$ is
the $T=0$ gap. Interestingly, the spectral weight is strongly
concentrated around some ``dispersion''
$E(Q)$\cite{Villain1975,James2009,Goetze2010}. 

The charge excitations in the $U>0$ Hubbard chain are in one-to-one
correspondence with the spin excitations of its $U<0$ counterpart via
a particle-hole transformation, such that the Mott charge gap
translates into a spin gap. Therefore, one may expect a Villain-like
mode to emerge in the dynamical {\it density-density} correlation
function in the repulsive case. Fig.\ref{fig:density} illustrates
this for $U=8$ and $T=1$, where we plot results for the operator $\hat
O(x)=(n(x)-1)$ ($n=n_\uparrow+n_\downarrow$ is the density operator). The offset
allows us to resolve the density fluctuations and eliminates large
contributions at $k=0$\cite{Nocera2016}. The low energies features can
be understood in the framework of the analytical strong-coupling
approach developed in Section \ref{sec:SC}.

\section{Analytic approach to the spin-incoherent regime at large $U$}
\label{sec:SC}
In this section we develop an analytic approach to the dynamics in the
spin incoherent regime in the $U\to\infty$ limit of the Hubbard model.
Our starting point is the Hamiltonian for the repulsive half-filled
Hubbard chain with open boundary conditions, Eq.~(\ref{Hubbard}).
The rationale for choosing open boundary conditions is that this will
make the following analysis simpler.

It is known from the exact solution, that the energy eigenvalues of the open
half-filled Hubbard model are equal to those of a tight-binding model of
spinless fermions and that the degeneracies in the spectrum are due to
the spin degrees of freedom\cite{EsslerBook}. Following
Ref.~\onlinecite{kumar09} we now construct a unitary transformation
to spinless fermion and spin degrees of freedom, that simplifies in
the $U\to\infty$ limit and reproduces these features.
We first express the fermion creation and annihilation operators in
terms of new spinless fermion operators $a_j$ and Pauli matrices
$\sigma^\alpha_j$ as
\bea
c^\dagger_{j,\uparrow}&=&\left[a^\dagger_j-(-1)^ja_j\right]\sigma^+_j\ ,\nn
c^\dagger_{j,\downarrow}&=&a^\dagger_j\frac{1-\sigma^z_j}{2}+(-1)^j
a_j\frac{1+\sigma^z_j}{2}\ .
\label{asigma}
\eea
The correspondence between the original spinful fermions and the spin
and spinless fermion degrees of freedom is\cite{kumar09}
\bea
|0\rangle_j&=&|-\rangle_j\otimes|\circ\rangle_j\ ,\
|\up\rangle_j=|+\rangle_j\otimes|\bullet\rangle_j\ ,\nn
|\down\rangle_j&=&|-\rangle_j\otimes|\bullet\rangle_j\ ,\
|\up\down\rangle_j=|+\rangle_j\otimes|\circ\rangle_j,
\eea
where $\sigma^z_j|\pm\rangle_j=\pm|\pm\rangle_j$ and
$a_j|\circ\rangle_j=0$, $|\bullet\rangle_j=a^\dagger_j|\circ\rangle_j$.
In terms of the operators \fr{asigma} the Hamiltonian reads
$H=H_\infty+H_1$, where
\bea
H_\infty&=&-t\sum_{j=1}^{L-1}P_{j}\left[a_j^\dagger
  a_{j+1}+{\rm h.c.}\right]
+\frac{U}{2}\sum_{j=1}^L\left[\frac{1}{2}-a^\dagger_j a_j\right],\nn
H_1&-&t\sum_{j=1}^{L-1}(-1)^j
\left[a_j^\dagger  a^\dagger_{j+1}+{\rm h.c.}\right]
\frac{\boldsymbol{\sigma}_j\cdot\boldsymbol{\sigma}_{j+1}-1}{2}.
\label{H}
\eea
Here we have defined
\be
P_{j}=\frac{1+\vec{\sigma}_j\cdot\vec{\sigma}_{j+1}}{2}.
\ee
The pairing term $H_1$ changes the number of doubly occupied sites,
which incurs a very large energy cost for $U\gg t$. As we are
interested in the infinite-$U$ limit we carry out a Schrieffer-Wolf
transformation to remove the pairing term in \fr{H} 
\be
e^{iS} H e^{-iS}=H_\infty+{\cal O}\Big(\frac{t^2}{U}\Big).
\label{remove}
\ee
Here the generator $S$ is taken in the form of a $t/U$-expansion 
\cite{macdonald88}
\be
S=\sum_{n=1}\left(\frac{t}{U}\right)^nS^{(n)}.
\ee
In order to obtain \fr{remove} we choose the leading term to be
\be
S^{(1)}=-i\sum_{j=1}^{L-1}(-1)^j
\left[a_j^\dagger  a^\dagger_{j+1}-{\rm h.c.}\right]
\frac{\boldsymbol{\sigma}_j\cdot\boldsymbol{\sigma}_{j+1}-1}{2}.
\ee
The main utility of the representation \fr{remove} is that the spin
degrees of freedom can now be removed from the Hamiltonian by a unitary
transformation\cite{kumar09}
\bea
{\cal U}&=&\prod_{\ell=1}^{L-1}\left[
(1-n_{\ell+1})+n_{\ell+1}P_{\ell}P_{\ell-1}\dots P_{1}\right].
\eea
This step makes use of the open boundary conditions and is more
involved in the periodic case. One has
\bea
\tilde{H}_\infty&=&{\cal U}^\dagger H_\infty {\cal U}\nn
&=&-t\sum_{j=1}^{L-1}\left[a_j^\dagger a_{j+1}+{\rm h.c.}\right]
-\frac{U}{2}\left[\hat{N}-\frac{L}{2}\right],
\label{Hinfty}
\eea
where $\hat{N}=\sum_{j=1}^La^\dagger_j a_j$. The Hamiltonian
\fr{Hinfty} is straightforwardly diagonalized by going to Fourier space
\be
a(k_n)=\sqrt{\frac{2}{L+1}}\sum_{j=1}^L\sin(k_n j)\ a_j,
\ee
where 
\be
k_n=\frac{\pi n}{L+1}\ ,\quad n=1,\dots L.
\label{quant}
\ee
In terms of the Fourier modes the Hamiltonian is diagonal
\bea
\tilde{H}_\infty&=&\sum_{k_n}\epsilon(k_n)a^\dagger(k_n)
a(k_n)+\frac{UL}{4},
\label{Hinfty2}
\eea
where the single particle dispersion is
\be
\epsilon(p)=-\frac{U}{2}-2t\cos(p)\ .
\ee

This shows, in accordance with the exact solution\cite{EsslerBook},
that in the infinite-$U$ limit the dynamics of the half-filled open Hubbard
model is determined by the non-interacting spinless fermion
Hamiltonian \fr{Hinfty}. For future reference we note that the ground
state of \fr{Hinfty} corresponds to a completely filled band 
\be
|{\rm GS}\rangle=\prod_{j=1}^La^\dagger_j|0\rangle\ ,
\label{GS}
\ee
We now want to use the strong-coupling formalism developed above to
determine dynamical correlation functions at finite temperatures. For
simplicity we begin by considering the density-density correlator.

\subsection{Density correlations at finite temperature}
\begin{figure}[ht]
\begin{center}
\epsfxsize=0.4\textwidth
\epsfbox{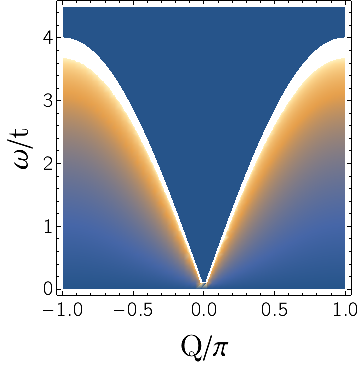}\qquad
\end{center}
\caption{Density-density correlator at low frequencies for $U/t=20$
and $\beta t=2$. The spectral weight increases with temperature in our
window \fr{window} and is concentrated around the ``dispersion''
\fr{dispersion}.} 
\label{fig:chi}
\end{figure}

We can use the above setup for determining density correlations in a
particular parameter regime of the half-filled Hubbard model as
follows. For strong interactions the Mott gap is proportional to the
Hubbard interaction $U$, while the characteristic energy scale in the
spin sector of the theory is $t^2/U$. The ``charge-sector only''
theory then applies (as the leading approximation in $t/U$) in the window
\be
\frac{t^2}{U}\ll T,\omega,t\ .
\label{window}
\ee
At $T=0$ the density-density response function vanishes for
frequencies below twice the Mott gap, but at $T>0$ a non-zero response
develops and can be determined in the framework of the model \fr{Hinfty}.
As the density of double occupied sites is very small in the regime
\fr{window}, we may apply the low-density approach of
Refs~\onlinecite{James2008,James2009,Essler2009,Goetze2010} to
finite-temperature dynamical correlation functions. The correlator of
interest is 
\bea
C_{\ell,m}(t)&=&\frac{1}{Z}{\rm Tr}\left(e^{-\beta
  H}[\rho_\ell(t),\rho_m(0)]\right)\ ,
\label{DSF0}
\eea
where
\be
\rho_j=\sum_{\sigma=\up,\down}c^\dagger_{j,\sigma}c_{j,\sigma}=1+\sigma^z_j(1-a^\dagger_ja_j)\ .
\ee
In the transformed basis we have
\bea
\tilde{\rho}_\ell&=&{\cal U}^\dagger\rho_\ell{\cal U}+{\cal O}(t/U)\nn
&=&1+(1-n_\ell)\sum_{j=0}^{L-\ell}\sigma^z_{L-j}\ P^{(\ell+1)}_j+{\cal O}(t/U),
\label{rhol}
\eea
where $P^{(\ell+1)}_k$ projects onto states with $k$ unoccupied sites
(in the spinless fermion variables) in the interval $[\ell+1,L]$ 
\be
P^{(\ell+1)}_k=\sum_{\ell<p_1<\dots<p_k\leq L}\prod_{j=1}^k (1-n_{p_j})
\prod_{\ontop{l=\ell+1}{l \neq p_s}}^Ln_l\ .
\ee
Here we have defined $P^{(L+1)}_0=1$. The leading contribution to
\fr{DSF0} in the framework of a $t/U$-expansion is given by
\bea
C_{\ell,m}(t)&\simeq&\frac{1}{Z}{\rm Tr}\left(e^{-\beta
  \tilde{H}_\infty}[\tilde\rho_\ell(t),\tilde\rho_m(0)]\right)\ ,
\label{DSF}
\eea
where 
\be
\tilde\rho_\ell(t)=e^{i\tilde{H}_\infty t}\tilde{\rho}_\ell
e^{-i\tilde{H}_\infty t}.
\ee
We note that at zero temperature the dynamical correlator \fr{DSF} is
of order ${\cal O}(t^2/U^2)$, see e.g. Ref.~\onlinecite{penc97}. This
is beyond the accuracy in $t/U$ we are working in here. The trace in
\fr{DSF} is over both charge and spin degrees of freedom. As these are
uncoupled in the leading order in the $t/U$-expansion we have
\be
\frac{1}{Z}{\rm Tr}\left(e^{-\beta  \tilde{H}_\infty}
{\cal O}_c{\cal O}_s\right)=
\frac{{\rm Tr}_c\left(e^{-\beta  \tilde{H}_\infty}{\cal O}_c\right)}
{Z_c}\frac{{\rm Tr}_s\left({\cal O}_s\right)}{2^L}\ ,
\ee
where ${\cal O}_{c,s}$ are any operators that act non-trivially only on
the charge and spin degrees of freedom respectively. As the spin
degrees of freedom do not have any dynamics at leading order in the
$t/U$-expansion we have
\be
\sigma^z_\ell(t)=\sigma^z_\ell\ ,\quad
2^{-L}{\rm Tr}_s\big(\sigma^z_\ell\ \sigma^z_n\big)=\delta_{\ell,n}\ .
\ee
This allows us to reduce the calculation of \fr{DSF} to the charge sector
only
\be
C_{\ell,m}(t)\simeq\sum_{j=0}^{L-{\rm min}(\ell,m)}
\frac{1}{Z_c}{\rm Tr}_c\Big[ e^{-\beta\tilde{H}_\infty}[{\cal
    A}_{\ell,j}(t),{\cal A}_{m,j}]\Big],
\label{C}
\ee
where
\be
{\cal A}_{\ell,j}(t)=\big(1-n_\ell(t)\big)P^{(\ell+1)}_j(t).
\ee
As the charge sector is a free theory \fr{C} can now in principle be
calculated using Wick's theorem and a determinant representation can
be obtained. As this is somewhat involved we focus on low temperatures
and proceed by applying the formalism of
Refs~\onlinecite{James2008,James2009,Essler2009,Goetze2010}, which is
possible as in the window \fr{window} the density of thermally excited
spinless fermions is small. To proceed, we express \fr{C} in a Lehmann
representation and then cast it in the form of a linked cluster
expansion. The leading contribution in our window \fr{window} is 
\be
C_{\ell,n}(t)\simeq\sum_{p}e^{\beta\epsilon(p)}
\langle p|[{\cal A}_{\ell,0}(t),{\cal A}_{n,0}]|p\rangle
+{\cal O}\left(e^{-\beta U}\right).
\label{leading}
\ee
Here $|p\rangle$ is a one-hole state 
\be
|p\rangle=\sqrt{\frac{2}{L+1}}\sum_{j}\sin(pj)\ a_{j}|{\rm GS}\rangle\ ,
\ee
with energy $-\epsilon(p)$ relative to the ground state, and
the momenta are quantized according to \fr{quant}. We note that
one-hole states have to be taken into account as we are working in 
the grand canonical ensemble, where the total number of spinless
fermions is not fixed. Evaluating the expectation value in
\fr{leading} gives 
\bea
C_{\ell,n}(t)&\simeq&\frac{4i}{L+1}{\rm Im}\sum_{p}e^{(\beta-it)\eps(p)}
\sin(p\ell)\sin(pn)\nn
&&\qquad\qquad\qquad\times\ G(\ell,n,t)\ ,
\eea
where $G(\ell,n,t)$ is the zero temperature Green's function of our
spinless fermions
\be
G(\ell,n,t)=\frac{2}{L+1}\sum_k \sin(k \ell)\sin(k n)\ e^{i\eps(k)t}.
\label{GF}
\ee
We now focus on $\ell,n$ in the middle of our open chain, where the
correlation functions are translationally invariant for sufficiently
large system sizes, so that $C_{\ell,n}(t)\longrightarrow {\cal
  C}(\ell-n,t)$. Taking the $L\to\infty$ limit we have
\be
{\cal C}(m,t)=2i{\rm Im} \int_{-\pi}^\pi\frac{dk\ dp}{(2\pi)^2}
e^{\beta\eps(p)+i(p+k)m+it\big(\eps(k)-\eps(p)\big)}.
\ee
We are interested in the real part of the Fourier transform of
${\cal C}(m,t)$
\bea
\chi(\omega,Q)&=&{\rm Re}\left[
\int_0^\infty dt\sum_{m=-\infty}^\infty e^{i\omega t-iQm}\ {\cal C}(m,t)
\right]\nn
&\simeq&\frac{1-e^{-\beta\omega}}{2}\theta(|4t\sin(Q/2)|-|\omega|\big)\nn
&&\times\
\frac{e^{\beta\epsilon(P+Q/2)}+e^{\beta\epsilon(P-Q/2-\pi)}}
{\sqrt{\big(4t\sin(Q/2)\big)^2-\omega^2}}\ ,
\label{chi}
\eea
where $\theta(x)$ is the Heaviside function and $P$ is given by
\be
P={\rm arcsin}\left(\frac{\omega}{4t\sin(Q/2)}\right)\ .
\ee
We see that in the window \fr{window} a finite-temperature resonance
develops as the temperature is increased, which follows the ``dispersion''
\be
\omega_Q=4t|\sin(Q/2)|.
\label{dispersion}
\ee
This is very reminiscent of the ``Villain-mode'' in the spin-1/2
Heisenberg chain\cite{Villain1975,James2009}. We note that the square
root singularity in \fr{chi} is an artefact of the low-density
expansion and will be smoothed by higher order
contributions\cite{Goetze2010,James2009}. In Fig.~\ref{fig:chi} we
plot $\chi(\omega,Q)$ for $U/t=20$ and $\beta t=2$. 
The results of the strong-coupling expansion can be compared to those
obtained by tDMRG, \emph{c.f.} Fig.~\ref{fig:density}. As the
numerical simulations were carried out at intermediate interaction
strengths $U=8t$ only a qualitative comparison is possible. The
dominant feature in the charge dynamical structure factor at $U=8t$
occurs at frequencies above approximately twice the $T=0$ Mott gap. 
This feature evolves smoothly with temperature and as we have remarked
before corresponds to the ${\cal O}(t^2/U^2)$ contribution in the
strong-coupling expansion. This is beyond the accuracy of the
analytical calculation presented above. At low frequencies within the
$T=0$ Mott gap the charge dynamical structure factor shown in
Fig.~\ref{fig:density} displays a weak feature, that is qualitatively
quite similar to the Villain-like mode obtained in the framework of
the strong coupling expansion, \emph{c.f.} Fig.~\ref{fig:chi}.

\subsection{Single-particle spectral function}
\begin{figure}[ht]
\begin{center}
\epsfxsize=0.4\textwidth
\epsfbox{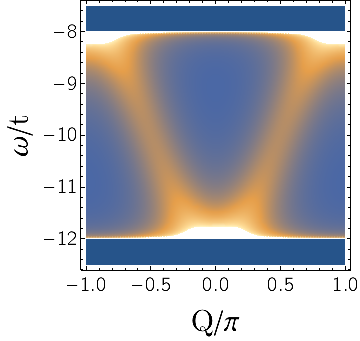}
\end{center}
\caption{Hole part of the spectral function at U=20t.}
\label{fig:a}
\end{figure}

We now turn to the single-particle Green's function
\be
{\cal G}(m,\ell,t)=\frac{1}{Z}{\rm Tr}\left[e^{-\beta H_\infty}\{
c_{m,\uparrow},  c^\dagger_{\ell,\uparrow}(t)\}\right].
\ee
At leading order in $t/U$ this becomes
\be
{\cal G}(m,\ell,t)\simeq\frac{1}{Z_cZ_s}{\rm Tr}\left[e^{-\beta \tilde{H}_\infty}\{
\tilde{c}_{m,\uparrow},  \tilde{c}^\dagger_{\ell,\uparrow}(t)\}\right],
\ee
where
\be
\tilde{c}_{m,\uparrow}={\cal U}^\dagger
{c}_{m,\uparrow}{\cal U}.
\ee
A complication compared to the case of the density operator considered
above is that the fermion operators are very complicated in the transformed
basis. However, it is possible to isolate the terms required for the
purposes of a low-temperature expansion. We find for $1<\ell<L$
\bea
\tilde{c}^\dagger_{\ell,\uparrow}&=&a^\dagger_\ell\prod_{j=\ell+1}^Ln_j
\sigma^+_{L-\ell+1} {\cal S}^\dagger_{L-\ell+1,L}\nn
&-&(-1)^\ell a_\ell\prod_{j=\ell+1}^Ln_j\
{\cal S}_{L-\ell+1,L}\ \sigma^+_{L-\ell+1}+\dots ,
\eea
where we have defined
\be
{\cal S}_{n,L}|\sigma_1,\dots,\sigma_L\rangle=
|\sigma_1,\dots,\sigma_{n-1},\sigma_{n+1},\dots
,\sigma_{L},\sigma_n\rangle\ .
\ee
The leading term in the low-temperature expansion of the single-particle
Green's function is then
\bea
{\cal G}(m,\ell,t)&\approx&(-1)^{\ell-m}
\big[G(\ell,m,t)+G(m,\ell,-t)\big]\nn
&\times&\frac{{\rm Tr}_s\left[\sigma^-_{L-m+1}{\cal S}^\dagger_{L-m+1,L}
{\cal S}_{L-\ell+1,L}\sigma^+_{L-\ell+1}\right]}{2^L}\nn
&=&\frac{(-1)^{\ell-m}\big[G(\ell,m,t)+G(m,\ell,-t)\big]}{2^{|\ell-m|+1}},
\eea
where $G(m,\ell,t)$ is given by \fr{GF}. Focussing again on the centre
of the chain where the Green's function is translationally invariant,
Fourier transforming and then taking the real part we obtain the
single-particle spectral function 
\bea
A(\omega,Q)&\approx&\frac{3\theta\big(2t-|\omega+U/2|\big)}{8t|\sin(k_0)|}\Big[
\frac{1}{5+4\cos(Q-k_0)}\nn
&&+\frac{1}{5+4\cos(Q+k_0)}\Big]+
\{\omega\rightarrow-\omega \}\ ,\nn
k_0&=&{\arccos}\Big(-\frac{\omega+U/2}{2t}\Big)\ .
\label{A}
\eea
The expression \fr{A} has threshold singularities at 
\be
|\omega\pm U/2|= 2t.
\ee
These are an artefact of working at lowest order in the low density
expansion and a summation of higher order contributions will smooth
out these singularities. In Fig.~\ref{fig:a} we plot $A(\omega,Q)$
for $U/t=20$.

The results of the strong-coupling low-temperature expansion can again
be compared qualitatively to those obtained by tDMRG. The hole part
of the spectral function for $U=8t$, $T=t$ shown in
Fig.~\fr{fig:U4-6}(c) looks very similar to the result displayed in
Fig.~\ref{fig:a}. The main differences are due to the finite bandwidth
of the spin degrees of freedom at $U=8t$, which leads to a bending of
the straight-line features at $\omega=-U/2\pm 2t$ in the
strong-coupling expansion. The general 
``$\overline{\underline{\rm V}}$-shaped'' structure of the response and
the concentration of spectral weight around the points
$\omega=-U/2-2t,\ Q=0$ and $\omega=-U/2+2t,\ Q=\pm\pi$ is already
clearly visible in the $U=8t$ data. 

\section{Conclusions}
\label{conclusions}
We have investigated the finite temperature behavior of the
photoemission spectrum of one-dimensional Mott insulators with
spin-charge separation. We considered the one-dimensional half-filled
Hubbard chain and determined dynamical response functions by a
combination of finite-temperature tDMRG methods and an analytic
strong-coupling approach in the spin-incoherent regime. The
single-particle spectral function displays an interesting
evolution with temperature. The spinon and holon branches in the
photoemission spectrum evolve into a broad dispersive band, while at
higher frequencies a second broad band of excitation
emerges. At elevated temperatures these features can be described in
terms of a picture reminiscent of the Hubbard-I approximation, where
spin correlations are completely washed out. Interestingly, unlike the
two-dimensional counterpart, in one-dimension the hopping amplitude
(effective mass) is not renormalized. In the window $4t^2/U\ll T\ll U$
a spin incoherent regime emerges. This can be described by an analytic
strong coupling expansion. The single-particle spectral function in
this regime displays a characteristic ``$\overline{\underline{\rm
    V}}$-shaped'' structure. 

In the density-density correlation function the ``melting'' of the
Mott gap is accompanied by the emergence of a temperature induced
resonance at low frequencies. This feature can be understood in terms
of transitions between thermally occupied levels and is akin to the
celebrated ``Villain-mode'' in gapped antiferromagnets. 


\acknowledgments
We thank R. Hulet and B. Bertini for illuminating
discussions. This work was supported by the U.S. Department of Energy,
Office of Basic Energy Sciences (AN and AEF) under grant
No. DE-SC0014407 (AEF) and the EPSRC under grant EP/N01930X/1 (FHLE).


\end{document}